# Anti-proximity Effect in Aluminum Nanowires

*Meenakshi Singh*[*‡], *Jian Wang*[*‡], *Mingliang Tian*[*‡], *T. E. Mallouk*[*†‡], *Moses H. W. Chan*[*‡]

[*] Department of Physics, Pennsylvania State University

[†] Department of Chemistry, Pennsylvania State University

[‡] Center for Nanoscale Science, Pennsylvania State University

ABSTRACT The anti-proximity effect, where the superconductivity in superconducting nanowires is suppressed or weakened when contacted by bulk superconducting electrodes, first revealed in arrays of Zn nanowires by tuning the electrodes from the superconducting to the normal state by means of an external magnetic field, has been confirmed in single crystal Aluminum nanowires. The critical current at zero magnetic field of an individual aluminum nanowire contacted by superconducting electrodes was found to be significantly smaller than that with normal electrodes showing that the effect is not a consequence of the magnetic field.

When a superconductor is placed in contact with a normal metal, signs of superconductivity appear in the normal metal. This 'proximity effect' is a much documented and well-studied phenomenon.[1] However, a number of recent experiments have reported an unexpected 'anti-proximity effect' (APE) in zinc nanowires (ZnNWs) contacted with bulk superconducting electrodes.[2-5] In the original experiment,[2] ZnNWs 6 µm in length embedded in track-etched polycarbonate membranes were squeezed between superconducting bulk electrodes for a quasi-4-electrode transport measurements. In 70 nm diameter nanowires contacted with bulk tin (Sn) electrodes, the superconducting transitions of Sn at 3.7 K and of the ZnNW at 1 K were both seen, consistent with expectation. However in 40 nm diameter nanowires, the transition for Sn was seen but the superconducting transition for Zn was absent. When the Sn electrodes were driven normal by a magnetic field of 300 Oe, the superconducting transition of Zn reappeared. The APE in the 40 nm Zn wires was replicated with Indium (In) and Lead (Pb) electrodes.[3] The experiment with In electrodes showed that the APE in the 40 nm Zn wires was switched off precisely when the magnetic field is increased above the critical field of In.[2] With Pb electrodes, the strength of APE was 'weaker' and there is no obvious 'suppression' of superconductivity in the ZnNWs in resistance measurements as a function of temperature or magnetic field. The APE however, showed up in the critical current ($I_c$) of the Zn wires. Specifically the $I_c$ of a Pb/Zn/Pb sample showed a dramatic increase when the magnetic field was increased towards the critical field ($H_c^{Pb}$) of Pb. Only at fields higher than $H_c^{Pb}$ the $I_c$ of the ZnNWs showed the 'normal' behavior, namely decreases with field (Figure 1). Similar $I_c$ behavior vs. magnetic field was also confirmed in four electrode measurements on e-beam assisted evaporated granular ZnNWs.[4,5] The nanowires and the electrodes for this experiment were evaporated in a single step using an e-beam lithographically fabricated mask. The authors attributed the increase in $I_c$ to the creation of quasi-particles as the bulk electrodes are driven toward the normal state by the external magnetic field. Since all the experiments to date make use of an external magnetic field to reveal the phenomenon, it is natural to wonder if the APE can be seen with no magnetic field.

This paper reports the observation of APE in aluminum nanowires (ANW) embedded in porous membrane and also in an individual single crystal ANW. The observation of the APE in a different superconducting nanowire with different electrodes suggests the universality of the phenomenon and in the case of the individual wire, the APE is indeed found without the aid of an external magnetic field.

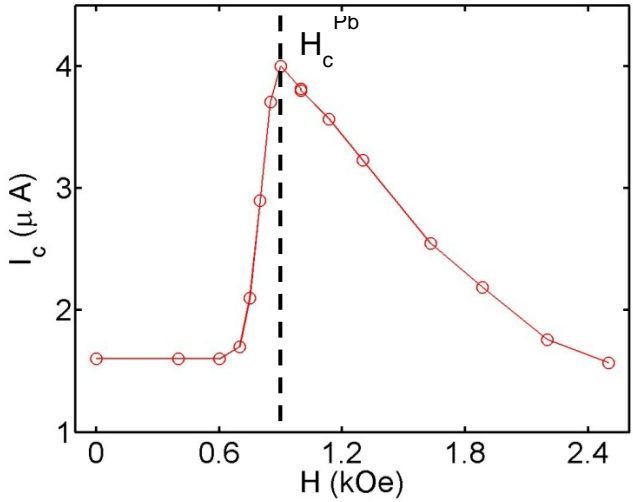

**Figure 1.** An enhancement in the critical current ($I_c$) of a Zn nanowire array is seen when the magnetic field is increased toward the critical field of the bulk superconducting Pb electrodes contacting the nanowires. The critical current peaks at the critical field $H_c^{Pb}$ of the electrodes. This graph is reproduced from reference 3.

Crystalline ANWs are synthesized by electrochemical deposition into the pores of an anodized aluminum oxide (AAO) membrane.[6] Characterization studies with X-ray diffraction, transmission electron microscopy and electron diffraction show the wires to be single crystal (high resolution TEM image can be seen in the inset of Figure 2(a)). A stable oxide layer 5 nm in diameter surrounds the nanowires, protecting them from uncontrolled oxidation. This resistance to uncontrolled oxidation makes these wires amenable to extraction from the AAO membrane for four probe measurements on an individual wire. The crystalline nature of these nanowires makes them good candidates for separating the effects of defect and morphology from intrinsic nanoscale physics. We report here the observation of the APE in single crystalline ANWs in two different sets of experiments. In the first set of experiments, transport measurements are made on crystalline ANWs of diameter 80 nm still embedded inside porous anodized aluminum oxide (AAO) membrane.[6] This experimental configuration is similar to that employed in references 2 and 3. For the second set of experiments, transport measurements on an individual ANW contacted by superconducting and normal electrodes were determined. Both superconducting (W) and non-superconducting (Pt) contacts were used in this experiment to highlight the effect of the superconducting electrodes. These transport measurements are performed on the nanowires using a Physical Properties Measurement System (Quantum Design Inc.), equipped with a Dilution Refrigerator (DR) or a $He^3$ Refrigerator (HR) and a superconducting magnet.

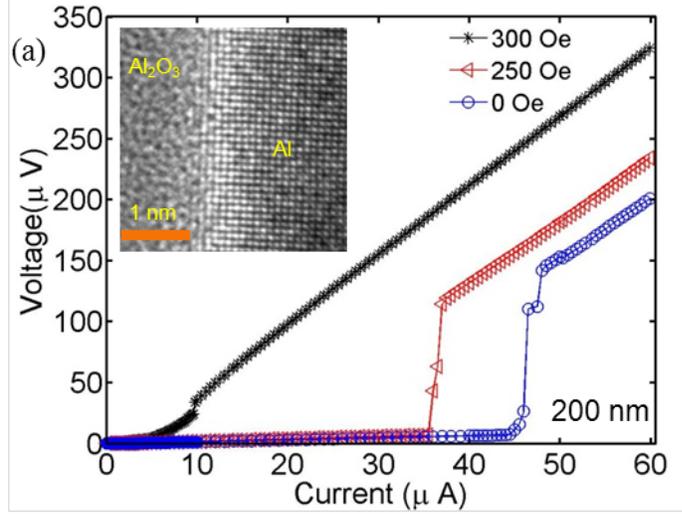

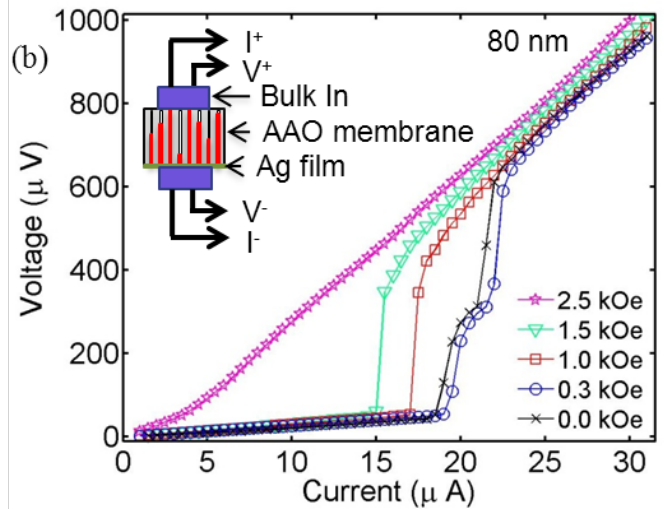

**Figure 2.** The inset of panel (a) is a high resolution TEM image of an Al nanowire, The oxidation layer and the crystalline nature of the nanowire can be seen. Inset of panel (b) shows the schematics of the measurement. The Al nanowires are embedded in an AAO membrane that is squeezed between bulk In electrodes. On one side the In is squeezed on a Ag film. Panel (a) shows the voltage vs applied current for a 200 nm diameter, 50 μm long Al naowire array measured at 0.1 K. The critical current $I_c$ decreases on the application of an external magnetic field. Panel (b) shows the corresponding measurement for an 80 nm diameter, 50 μm long Al nanowire array at 0.4 K. The $I_c$ increases compared to the 0 Oe value when a 300 Oe field is applied.

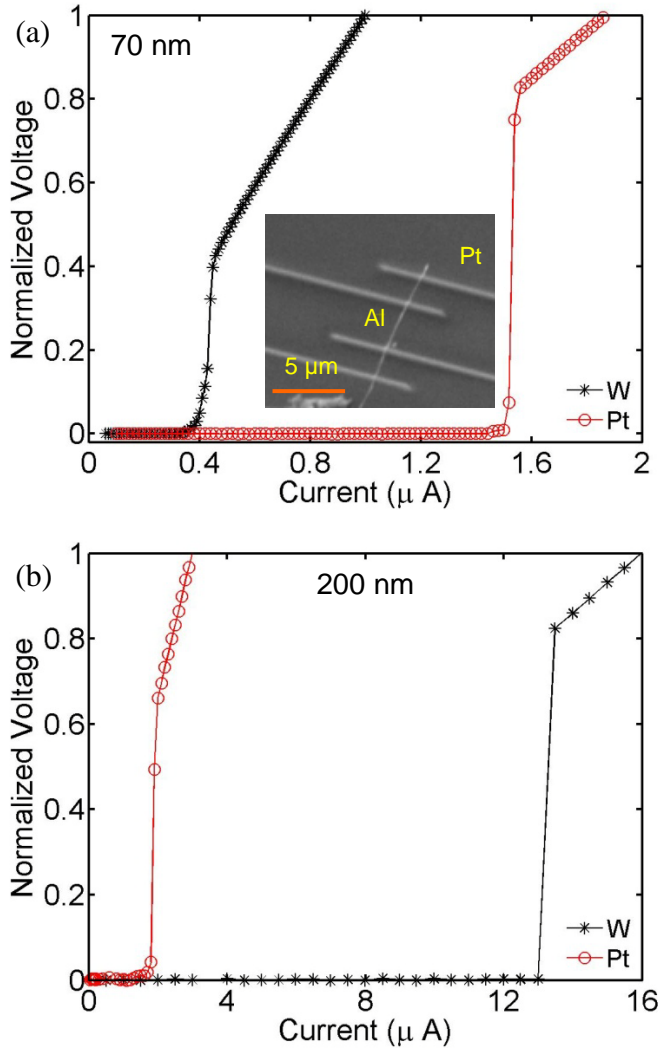

**Figure 3.** Panel (a) shows normalized voltage vs. applied current for two 70 nm diameter, 2.5 μm long single ANWs at 0.1 K. One of the nanowires is measured using normal Pt electrodes and the other is measured using superconducting W electrodes. The inset shows a scanning electron micrograph of the ANW contacted with the Pt electrodes. Panel (b) shows normalized voltage vs. applied current for two 200 nm diameter, 4 μm long single ANWs. The critical current with the superconducting W electrodes is higher than the critical current with the normal Pt electrodes.

Figure 2(a) shows Voltage (V) vs. excitation current (I) measurements at different applied fields (H) of an array of 50 μm long ANWs of 200 nm diameter still embedded in the membrane at 0.1 K. The wires are directly in contact with a bulk In dot of 1 mm in diameter on one side of the membrane. On the other side, the bulk In electrode is squeezed onto an evaporated Ag film (250 nm thick) covering the membrane. The geometry of the measurement, with the applied magnetic field parallel to the nanowires is shown in the inset of Figure 2(b). In the absence of a magnetic field, when the In electrode is superconducting, the $I_c$ of the nanowires is 47 μA. An externally applied magnetic field of 300 Oe (the critical field for bulk In is 300 Oe) drives the In electrode normal, as seen by the non zero resistance at low currents, the $I_c$ of the nanowires decreases. This decrease in the $I_c$ of the nanowires in the presence of an external magnetic field is consonant with expectations. When the same experiment is performed on a 80 nm diameter nanowire array however, the behavior of $I_c$ is different. The V vs. I measurements are shown in Figure 2(b). It is estimated that 15 nanowires are being contacted. The non-zero residual resistance At 0 Oe field may be from the Ag film coating the membrane. The $I_c$ of the nanowires when the In electrodes are superconducting (O Oe field) is 21 μA. On driving the In electrode normal (300 Oe field)), the $I_c$ of the nanowires increases to 22 μA. The enhancement in the $I_c$, while only 1 uA, is significantly larger than the instrument error of 0.01 μA. This is a sign of a more robust superconductivity in the ANWs contacted by normal electrodes and a signature of the APE. In this and other similar samples, the $I_c$ of the nanowire increases on the application of a magnetic field and reaches a maximum at the $H_c$ of the In electrodes, much like the effect seen in Zn nanowires with Pb electrodes (Figure 1). The fact that APE is seen in thin but not in thicker wire is consistent with the findings in the ZnNW [2,3], where APE was seen in 40 nm wires but not in 70 nm wires. The difference in the 'characteristic' diameters determining the presence or absence of APE in ZnNWs and ANWs may be a consequence of the different materials and possibly also the different crystalline quality of the wires. It is also noteworthy that APE is seen in wires up to 50 μm in length. The APE has also been seen in Zn wires as short as 1 μm in length [5] indicating that the APE is present over a wide range of the length of the nanowire.

In the second set of measurements, crystalline ANWs were released by dissolving the alumina membrane and precipitating and dispersing the resultant solution on a silicon substrate with a 1 um thick $Si_3N_4$ insulating layer. The sample was then transferred into a focused ion beam deposition and etching system (FIB/SEM FEI Quanta 200 3D). An isolated nanowire was located using the electron microscope and four FIB-assisted Pt or W electrodes were deposited on it for a standard four electrode measurement.[7,8] During FIB-assisted deposition, the ion beam etches away the native oxide layer on the ANW to ensure good ohmic contact to the wire.[9,10] Since the oxide layer on the 80 nm diameter ANWs is 5 nm thick, the effective diameter of the single nanowires is 70 nm. FIB deposited Pt is normal but W is superconducting with a $T_c$ of 4.8 K and $H_c$ of 8 T.[11] The $H_c$ of a 70 nm ANW is ~ 5000 Oe. Measurements were made on two different single 70 nm diameter wires, one of which was contacted with four normal Pt electrodes and the other with four superconducting W electrodes (the inset of Figure 3(a) is a scanning electron micrograph showing the geometry of the measurement). The distance between the voltage electrodes for both wires is ~ 2.5 μm. Figure 3(a) shows that the $I_c$ of the wire with the normal Pt electrode is 1.5 μA while the $I_c$ of the wire with superconducting W electrodes is 0.4 μA at 0.1 K. An identical experiment was performed on 200 nm diameter ANW. The $I_c$ with normal Pt electrodes is 2 μA while the $I_c$ with the superconducting W electrodes is 13 μA (Figure 3(b)). The weakening of superconductivity in the presence of a bulk superconductor is again not seen in the thick 200 nm diameter ANWs. The $I_c$ for the 200 nm wire contacted by W electrodes is actually much higher than that with Pt electrodes. This may be a consequence of the standard proximity effect of the W electrodes on the thick Al wire. As there are no applied magnetic fields in this experiment, this serves to definitively decouple the APE seen

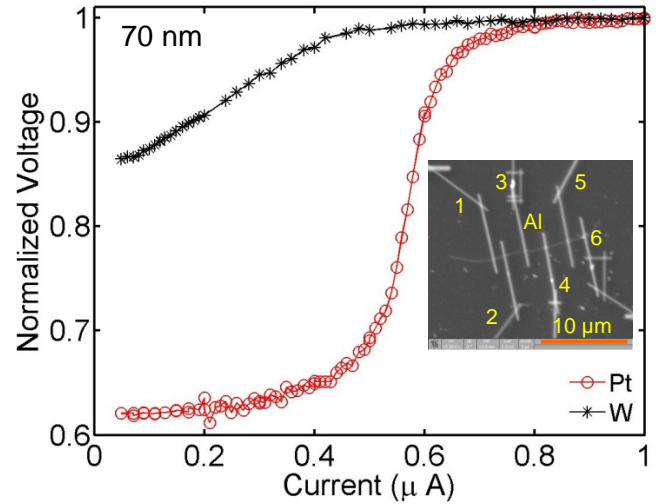

**Figure 4.** The inset of the figure shows an SEM image of the sample being measured. An 70 nm Al nanowire is contacted with 6 electrodes. Electrode 1, 2, 3 and 6 are made with normal Pt and electrodes 4 and 5 are made with superconducting W. Electrode 1 and 6 are used to pass current through the nanowire. Normalized voltage vs. applied current for segment 2-3 between normal Pt electrodes and segment 3-4 between one Pt and one superconducting W electrode is measured at 0.5 K. The length of both the segments is ~ 2 μm and the diameter of the wire is 70 nm. The $I_c$ of the wire segment between the Pt electrodes is higher

in the 70 nm wire from negative magnetoresistance caused by a weak applied magnetic field.[12-14] It is worth noting that even though the length of 70 nm diameter ANW between the W electrodes is only 2.5 μm, the superconductivity of the ANW is not completely suppressed by the W electrodes, unlike the case of the 2 μm long ZnNWs with Sn electrodes.[2] An enhancement of $I_c$ rather than a complete suppression of superconductivity by the superconducting electrode is more akin to the behavior of ZnNWs with Pb electrodes.[3] This points to the APE strength being dependent on the nature of the superconducting electrode.

With the aim of measuring $I_c$ of the same 70 nm diameter ANW with both superconducting and normal electrodes, a 6 electrode geometry was used (inset of Figure 4). Current was introduced using two normal Pt electrodes (#1, #6 in the inset of Figure 4) and two superconducting W (#4, #5) and two normal Pt (#2, #3) electrodes were deposited to measure the voltage. Graphs of the normalized voltage measured between two normal Pt electrodes #2 and #3 (L = 2 µm) and one normal Pt and one superconducting W electrodes #3 and #4 (L = 2 µm) measured at 0.5 K are shown in Figure 4. The source of the relatively large residual resistance may be the FIB process in which the ion beam might have damaged the area close to the electrodes and make it non-superconducting. The $I_c$ for the segment with both Pt electrodes is 0.45 µA whereas the $I_c$ for the segment with one Pt and one W electrodes is 0.25 µA. Another 70 nm ANW was measured in the 6-electrode configuration similar to that shown in the inset of Figure 4. The $I_c$ for the segment with both Pt electrodes is 0.46 µA whereas the $I_c$ for the segment with W electrodes is 0.1 µA. We do not know why the $I_c$s measured with the 6 electrode configuration are different from those shown in Figure 3(a). It is likely that with additional FIB processes the ANW is made thinner than the wires contacted with only Pt or only W electrodes. Nevertheless, this measurement confirms the results shown in Figure 3(a) namely, the superconductivity of the ANW segment close to the superconducting W electrode is weaker than the superconductivity of the Al NW near the normal Pt electrodes. The APE is seen here again in the absence of an external magnetic field, just by changing the nature of the electrodes.

A theoretical model of the APE was proposed by Fu *et al.*[15]. In their model the contacting electrodes are modeled as two resistors connected in series with the nanowire with a capacitor that is in parallel with the nanowire. When the electrodes are normal, the resistors provide a medium for dissipation of quantum phase slips, thus stabilizing the superconductivity. When the electrodes are superconducting, this dissipation path disappears and the wire is driven normal by the phase slip process. Although the model provides qualitative explanation of the experimental results, it does not explain how the strength of APE effect depends on the material of the contacting electrodes. In addition to the material of the electrodes, other important parameters in these experiments are the 'characteristic' diameter and the length of the nanowires that define the presence or absence of APE. It is difficult to obtain quantitative predictions for these quantities from this model.

A more quantitative mechanism using the time independent Ginzburg-Landau equations was proposed by Vodolazov *et al.*[16] This model is strictly applicable only near $T_c$, the superconducting transition of the nanowire. The authors expect the results obtained near $T_c$ also to be applicable at low temperatures. This mechanism of Vodolazov uses the fact that the diameters of the wires are smaller than the superconducting coherence length and model them as 1D systems. The coherence length for the ANWs in these experiments can be estimated using the value of $\rho.l$ (where $l$ is the mean free path) as $4 \times 10^{-16}$ Ω m$^2$,[17] the dirty limit coherence length $\xi(0) = 0.855(\xi_0 l)^{0.5}$ and 1600 nm as the value for $\xi_0$. $\xi(0)$ is estimated to be ~ 50 nm. In comparison, APE was seen in 70 and 80 nm diameter ANW but not in 200 nm diameter nanowires. This result is qualitatively consistent with the model. However, in ZnNWs, the $\xi(0)$ was estimated to be ~ 150 nm and the APE was seen in 40 nm diameter nanowires but, not in 70 nm diameter nanowires.[2] The model also predicts a weakening or absence of the antiproximity in nanowires with length L > $\Lambda_Q$ (the charge imbalance length). The charge imbalance lengths for the ANW and Zn NWs were calculated [18] to be ~ 19 µm and 22 µm respectively. The result again, is qualitatively consistent with the finding of a weak APE in a 50 µm long ANW and 30 um long Zn NWs. The weakness of the model lies in the fact that it proposes the enhancement of $I_c$ is a consequence of an external magnetic field. In this model the APE is understood via a charge imbalance created at the boundary of electrode and the nanowire. This boundary approaches the center of the nanowire as the applied magnetic field is increased and the effective length of the nanowire is shortened reducing the chances of phase slips and therefore increasing $I_c$. The result reported here, that an external magnetic field is not required for the observation of the APE, indicates a reformulation of the model is needed.

In conclusion, the APE has been seen in crystalline Al nanowires in a variety of geometries. The APE has also been seen in the absence of magnetic fields, clearly establishing that it is a function of the bulk measuring electrodes. Further experiments exploring this effect are underway.

ACKNOWLEDGMENT The authors acknowledge use of facilities at Materials Characterization Lab at Penn State University. This work was supported by the Center for Nanoscale Science (Penn State MRSEC) funded by NSF under grant no. DMR-0820404.